\documentclass[sigconf]{acmart}

\usepackage[english]{babel}
\usepackage{blindtext}

\acmYear{2018}
\acmConference{CoNEXT '18}{December 4--7, 2018}{Heraklion/Crete, Greece}

\usepackage{times,amsmath,epsfig}
\usepackage{textcomp}
\usepackage{multirow}
\usepackage{graphicx}
\usepackage{balance} 
\usepackage{booktabs}
\usepackage[ruled,vlined,linesnumbered]{algorithm2e}
\usepackage{comment}
\usepackage{subfigure}
\usepackage{subfloat}
\usepackage{color}
\usepackage{xcolor}
\usepackage{soul}
\usepackage{marginnote}

\theoremstyle{plain}

\definecolor{mycolor}{rgb}{0.77, 0.89, 0.99}

\begin{document}


\title{Effects of Packet Loss and Jitter on \\ VoLTE  Call Quality}
\author{Elena Cipressi$^{1,2}$ and Maria Luisa Merani$^1$}
\affiliation{
\institution{$^1$Dipartimento di Ingegneria ``Enzo Ferrari'', University of Modena and Reggio Emilia, Modena, Italy\\$^2$Empirix Inc., Modena, Italy}
}
\email{email: {elena.cipressi, marialuisa.merani}@unimore.it}

\begin{abstract} 
This work performs a preliminary, comparative analysis of the end-to-end quality guaranteed by Voice over LTE (VoLTE), examining  
several millions of VoLTE calls that employ two popular speech audio codecs, namely, Adaptive Multi-Rate (AMR) and Adaptive Multi-Rate WideBand (AMR-WB). To assess call quality, VQmon\textregistered , an enhanced version of the standardized E-Model, is utilized. The study reveals to what extent AMR-WB based calls are more robust against network impairments than their narrowband counterparts;
it further shows that the dependence of call quality on the packet loss rate is approximately exponential when the AMR codec is used, whereas it is nearly linear for the AMR-WB codec.

\keywords{Call Detail Record; Network Measurements; Speech Codecs; Voice Quality; VoLTE}
\end{abstract}
\settopmatter{printfolios=true}
\maketitle
\section{Introduction and Background}
\label{sec:Introduction}
VoLTE is a recent technology, allowing to perform voice calls in LTE cellular systems via the IP Multimedia Subsystem (IMS), an all-IP architectural framework integrated on top of the LTE core network. 
As for the majority of Voice over IP (VoIP) services, in VoLTE digital voice samples are properly placed in Real-Time Transport Protocol (RTP) packets  and constitute the data session of the voice call, whereas the signaling of the call is handled via the Session Initiation Protocol (SIP). 
Various codecs can be employed to produce the digital version of the speech:  unlike the Public Switched Telephone Network (PSTN), where the codec choice is limited to equipments covering voice components  within the $[300,3400]$ Hz frequency window, LTE can also leverage on wideband codecs, that span a wider frequency range, from $50$ to $7000$ Hz and higher frequencies; this guarantees an increased intelligibility and naturalness to the reconstructed speech, and ultimately an improved quality being experienced by VoLTE users. 
AMR \cite{3GPPamr} and AMR-WB \cite{3GPPamrwb} are among the most commonly deployed codecs in contemporary LTE networks, and as such are the subject of the current investigation.
As regards end-to-end voice quality assessment,
VQmon\textregistered ~\cite{Clark01modelingthe} is the objective, non-intrusive tool employed in this study; it is an extension of the well-established E-Model~\cite{ITU_EMODEL}, 
and exactly like the latter, it provides an output value between $0$ and $100$, the so-called Rating factor, R-factor for short, to grade the overall call quality. A suitable correspondence allows to map the R-factor to the more popular Mean Opinion Score (MOS) on a $1$ to $5$ scale. 
Such correspondence has been recently updated for the Wide-band version of the E-Model~\cite{ITU_WBEMODEL}, where the R-factor can reach values up to $129$, as it is the case for the AMR-WB codec. 
On the literature rim, numerous studies have been recently conducted to understand VoLTE behavior.
In \cite{Elnashar-Transvt2017}, the performance of VoLTE and of Circuit-Switched Fall Back was benchmarked, pinpointing what values of call set up delay can be achieved under various radio conditions.
In \cite{VoLTEcoverage},  the authors' objective was to understand whether the adoption of a lower bit rate of the AMR-WB codec could result in an augmented coverage for VoLTE users.
Differently from the previous contributions, the aim of this paper is to discern the dependency of VoLTE call quality on network impairments, i.e., packet loss and jitter, and to grasp the influence that different codec choices, namely, AMR or AMR-WB, have on end-to-end speech quality.
Accordingly, a significantly large set of VoLTE calls is examined: the network conditions they encountered were recorded and their quality estimated via VQmon\textregistered. 
The obtained results allow to realistically compare the behavior of AMR and WB-AMR codecs and to shed light on VoLTE performance.

\vspace{-0.07truecm}
\section{Setting}
\vspace{-0.07truecm}
We conducted this study on over ten millions of VoLTE calls, collected within a single commercial LTE network from an urban area, over the first half of 2018.
Several relevant information and metrics about the RTP voice flows were captured by a proprietary probe on the $Mb$ interface\cite{IMS}, anonymized and aggregated in a \textit{.csv} file. 
Positioning the tapping point at the $Mb$ interface allowed to collect call detail records for both directions, i.e., for the voice flow being generated by the calling party and for the flow originated by the callee.
We chose to analyze the worst among the two directions, i.e., the uplink, therefore capturing the negative effects that the Radio Access Network (RAN) traversal has on voice packets.
For each flow, the examined records were the number of transmitted packets, the number of received packets, the average and maximum jitter, $J_{max}$, the R-factor computed according to VQmon\textregistered, and the type of codec being used.
Moreover, a jitter buffer emulator (JBE) was instantiated, in order to realistically model the compensation that takes place receiver side, smoothing out the delay variations that voice packets exhibit after traversing the network. The emulator forced a delay on packets that arrived early, and immediately forwarded late packets.  
In our system, the JBE was set to receive initial packets with a $50$ ms delay, then to dynamically modify its play-out delay according to the average jitter of the previous $16$ packets.
Under these assumptions, we were able to estimate the packet loss rate, $P_{loss}$, evaluating the ratio of the number of lost/excessively delayed packets to the total number of received packets after the JBE. 
Filtering out
invalid data and neglecting the calls that either employ the Enhanced Voice Services (EVS) wideband codec \cite{EVS} or alternative, less popular speech codecs, we were left with $10,862,591$ voice flows. They were  further distinguished in AMR and AMR-WB based, amounting to $71\%$ and $29\%$, respectively. 

\vspace{-0.07truecm}
\section{Results and Discussion}
\begin{figure}[tbh]
	\vspace{-0.15truecm}
	\includegraphics[width=0.47\textwidth]{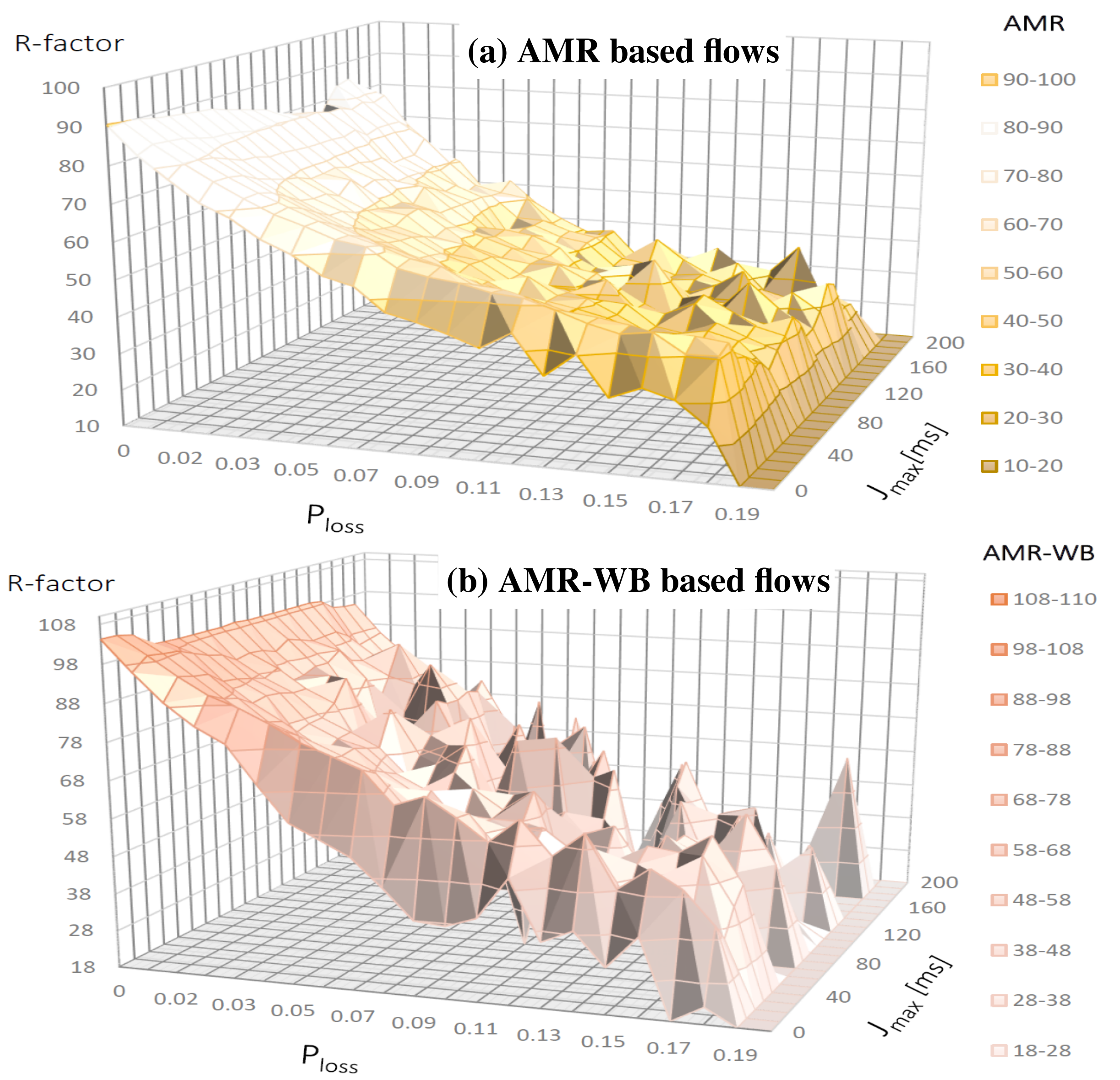}
	\vspace{-10pt}
	\caption{R-factor as a function of  $P_{loss}$ and $J_{max}$}
	\vspace{-5pt}
	\label{fig:R_factor_3D}
	\vspace{-2pt}
\end{figure}

\begin{figure}[htb]\centering
	\includegraphics[width=0.47\textwidth]{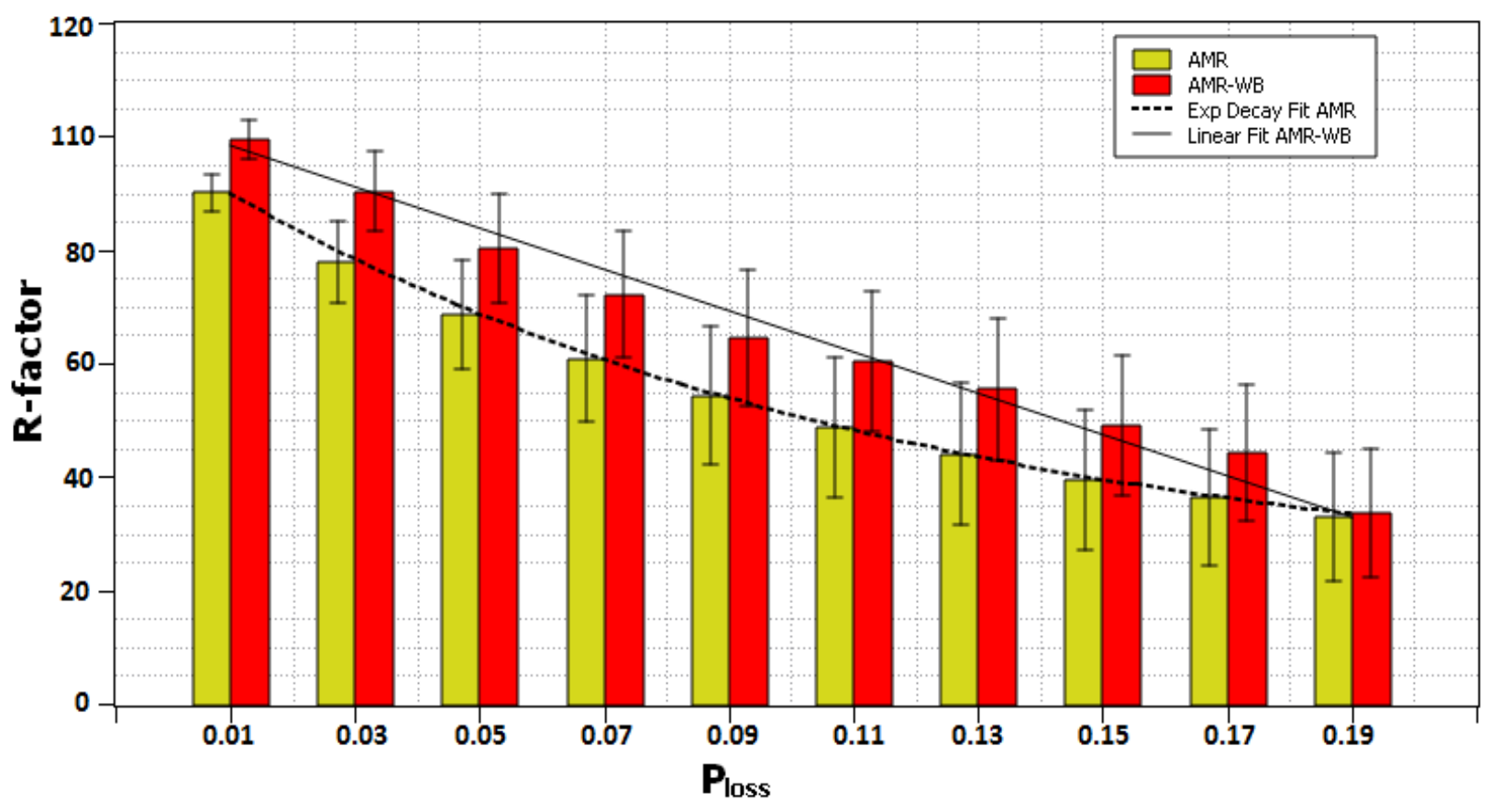}
	\vspace{-15pt}
	\caption{R-factor as a function of $P_{loss}$
	}
	\vspace{-12pt}
	\label{fig:PLavgAMRAMRWB}
	\vspace{-0.10truecm}
\end{figure}

\vspace{-0.07truecm}
 Figs.\ref{fig:R_factor_3D}(a) and \ref{fig:R_factor_3D}(b) display the R-factor of the examined flows as a function of $P_{loss}$ and $J_{max}$, for the AMR and AMR-WB case, respectively. The comparison between the two figures indicates  that the adoption of the AMR-WB codec guarantees higher R-factor values and leads to a less pronounced dependence of it on $P_{loss}$ and $J_{max}$. 
The jagged behavior appearing in Fig.\ref{fig:R_factor_3D}(b) is due to the lack of points in the region of high values of packet loss rate and maximum jitter; as a matter of fact, the examined LTE network is well designed, so that the values such network impairments exhibit are  modest.
Moreover, both figures demonstrate that it is the packet loss rate $P_{loss}$ to most significantly influence the end-to-end quality of VoLTE calls.
Next, Fig.~\ref{fig:PLavgAMRAMRWB} reports the average R-factor and its standard deviation values over $10$ uniform intervals of packet loss rate, when $P_{loss}$ varies between $0$ and  $0.2$.
\noindent It is interesting to observe the sharp decay that the average R-factor displays for the AMR case, whereas the decrease is less pronounced for the AMR-WB case. The standard deviation tends to increase for increasing values of the packet loss rate, but this has to be mainly ascribed to a decreasing size of the population of samples. 
For the AMR case, this figure also shows the first order, exponential  fit performed on the set of 
$(x_i,y_i)$ points, $i=1, 2, \ldots, 10$, where $x_i$ represents the median value of $P_{loss}$ in every interval and $y_i$  the value of the corresponding average R-factor.
We have used Levenberg-Marquardt algorithm, 
choosing  $y(x)=y_0+A e^{-x/B}$ as the fit function (dashed line). The $y_0$, $A$ and $B$ values are $17.953$, $71.63$ and $0.12$, respectively.
The fitting is truly satisfying, and leads to the conclusion that the quality dependence of AMR VoLTE calls on $P_{loss}$ replicates the Quality of Experience (QoE)  exponential dependence on the Quality of Service (QoS) parameters first outlined in \cite{IQX}.
This figure also reports the linear regression for the AMR-WB case, using $y(x)=A^{'} + B^{'}x$ as the fit function (solid line). 
The $A^{'}$ and $B^{'}$ values are $99.01$ and $-340.70$, respectively. 
The coefficient of determination that quantifies the fitting goodness (a.k.a. $R^{2}$) is equal to $0.98$, demonstrating that the linear fit is adequate. 
Although not reported on the figure, we verified that the exponential fit is not satisfying for the AMR case. 
\vspace{-0.07truecm}
\section{Conclusions}
\vspace{-0.07truecm}
Examining over ten million VoLTE calls, this study has demonstrated to what extent AMR-WB flows are more robust than AMR ones against packet losses and jitter.
Moreover, the analysis has revealed that the R-factor dependency on the packet loss rate is successfully captured by an exponential law for the calls performed via the AMR codec, whereas it follows a linear decay trend for the AMR-WB case.


\bibliographystyle{ACM-Reference-Format}
\bibliography{reference}

\end{document}